\documentclass[
twocolumn,
prl,showpacs,preprintnumbers,amsmath,amssymb]{revtex4}

\usepackage{graphicx}
\usepackage{amssymb}
\usepackage{amsmath}
\usepackage{dsfont}
\usepackage{bm}
\usepackage{mathrsfs}

\begin{document}
\title{Ambiguities in the scattering tomography for
central potentials}
\author{Awatif Hendi$^{1,2}$, 
Julian Henn$^2$,  
and Ulf Leonhardt$^2$}
\affiliation{
$^1$Science and Medical Studies,
Female Section,
King Saud University,
P.O. Box 22452, Riyadh 11495, Saudi Arabia\\
$^2$School of Physics and Astronomy, University of St Andrews,
North Haugh, St Andrews KY16 9SS, Scotland
}
\begin{abstract}
Invisibility devices 
exploit ambiguities in the inverse scattering problem
of light in media.
Scattering also serves as an important general tool 
to infer information about the structure of matter. 
We elucidate the nature of scattering ambiguities 
that arise in central potentials.
We show that 
scattering is a tomographic projection:
the integrated scattering angle is a projection of 
a scattering function onto the impact parameter.
This function depends on the potential, 
but may be multi-valued, allowing for ambiguities
where several potentials share the same
scattering data.
In addition, multivalued scattering angles 
also lead to ambiguities.
We apply our theory to show that it is 
in principle possible to construct
an invisibility device without infinite phase velocity of light.
\end{abstract}
\date{\today}
\pacs{42.79.-e, 03.65.Wj, 03.65.Nk,}
\maketitle

An invisibility device \cite{Gbur,Pendry,LeoConform,LeoNotes,Kerker,Others}
should guide light around an object as if nothing were there.
It is conceivable that
such devices can be made using modern metamaterials
\cite{Pendry,LeoConform,LeoNotes,Others}.
Passive optical devices use spatially varying refractive-index
profiles for imaging.
Within the validity range of geometrical optics,
index profiles of isotropic dielectric media 
are mathematically equivalent to potentials
for light rays \cite{BornWolf,LeoNotes}.
Therefore, such an invisibility device corresponds to a potential
that has the same scattering characteristic as empty space.
While the inverse scattering problem for waves has unique
solutions \cite{Nachman},
the scattering of rays may be ambiguous.
Here we show how such ambiguities arise in the case
of radially symmetric potentials.
Our theory indicates  that it is 
in principle possible to construct
an invisibility device where the phase velocity of light
does not approach infinity,
in contrast to all previous proposals
for macroscopic cloaking \cite{Pendry,LeoConform,LeoNotes}.
This could inspire ideas for developing 
invisibility devices without anomalous dispersion \cite{Pendry}
that could operate in a relatively wide frequency window.
In addition to applications in a potentially new area
for metamaterials,
our theory has wider implications for the field of scattering tomography.

The inversion of the classical scattering in central potentials
is a classic textbook problem
that has made it into the exercises in
Landau's and Lifshitz' {\it Mechanics} \cite{LLproblem}.
Since Rutherford's experiments,
scattering has served as an important tool to investigate
the structure of matter, with modern applications
ranging from biomedical research to astrophysics.
Techniques to infer the structure of matter from scattering
are often called scattering tomography,
although, strictly speaking, 
they are not directly related to traditional 
tomography \cite{Tomo} where the shape of a hidden
object is reconstructed from projections.
Here we show that the case of scattering in central potentials
literally is a tomographic projection in disguise, 
but with an interesting twist:
the object to be reconstructed corresponds to the potential,
but may be represented by a multi-valued function,
allowing for ambiguities.

\begin{figure}[t]
\begin{center}
\includegraphics[width=20.0pc]{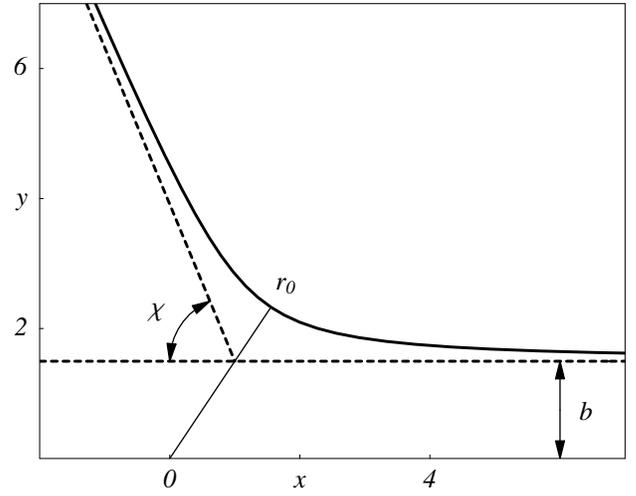}
\caption{\label{fig:scattering}
{\tiny
Scattering in a central potential.
A trajectory incident with impact parameter $b$ is deflected
by the angle $\chi$ in the rotationally symmetric
potential $U(r)$ centered at the origin,
with $r=\sqrt{x^2+y^2}$ in 
the Cartesian coordinates $x$ and $y$.
The turning point of the trajectory is denoted by $r_0$.
The figure shows Rutherford scattering \cite{LL1}
in a repulsive $1/r$ potential.}
\vspace*{-8mm}
}
\end{center}
\end{figure}

\begin{figure}[t]
\begin{center}
\includegraphics[width=19.0pc]{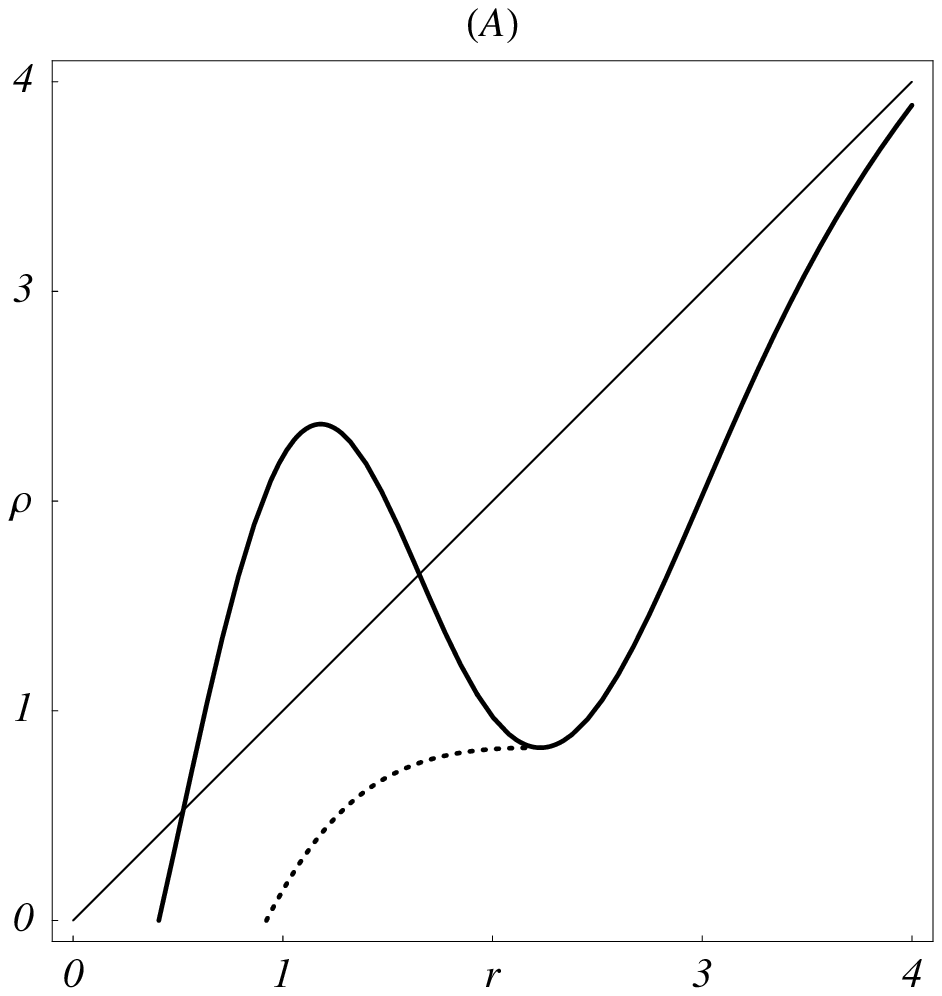}
\includegraphics[width=20.0pc]{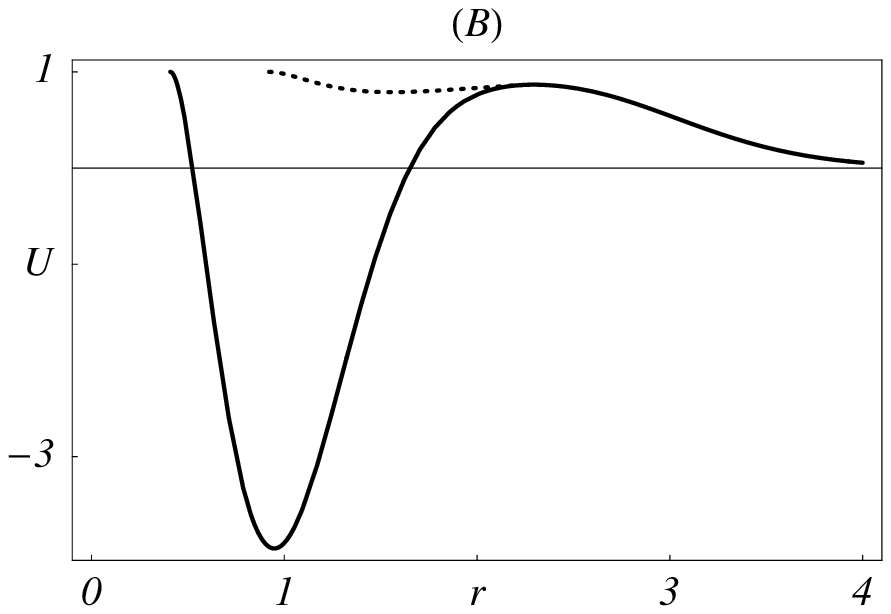}
\caption{\label{fig:rhohide}
{\tiny
Representation of the potential by the turning parameter
$\rho$ defined in Eq.\ (\ref{eq:rho}).
The solid line in A shows a potential with
a fold in the turning parameter where $r(\rho)$
is multi-valued. 
Subfigure B shows the corresponding potential $U(r)$.
The dotted lines describe $\rho(r)$ and $U(r)$
for a potential with the same scattering characteristics.
Here $r(\rho)$ 
was obtained from Eq.\ (\ref{eq:dw})
using the definition (\ref{eq:w}) of the scattering
function $W$.}
\vspace*{-10mm}
}
\end{center}
\end{figure}

Figure \ref{fig:scattering} illustrates the situation 
typical for scattering in central potentials.
An incident ray characterized by the impact parameter $b$
and the energy $E$ is deflected by the angle $\chi$.
We use polar coordinates with radius $r$
and angle $\varphi$
in the plane orthogonal 
to the angular-momentum vector.
The scattering angle is determined as \cite{LL1}
\begin{equation}
\chi = \pi - 2\int_{r_0}^\infty
\frac{(b/r)\,dr}{\sqrt{\rho^2-b^2}} \,.
\label{eq:chi}
\end{equation}
Here $r_0$ denotes the turning point of the trajectory
given by $b$ and $E$,
and $\rho$ represents the potential as
\begin{equation}
\rho= r\sqrt{1-\frac{U(r)}{E}} \,,\quad
\frac{U}{E} = 1 - \frac{\rho^2}{r^2} \,.
\label{eq:rho}
\end{equation}
The turning point is given by the largest value of $r$
at which the denominator in the integrand (\ref{eq:chi})
of the scattering angle vanishes,
{\it i.e.} at which $\rho=b$.
Reversing this relation leads to 
a physical interpretation for $\rho$:
$\rho(r)$ describes the impact parameter for which
the radius $r$ is a turning point.
Therefore we may call $\rho$ turning parameter.
Figure \ref{fig:rhohide} illustrates the representation
of the potential using the turning parameter.
The largest zero of $\rho(r)$ corresponds
to the potential barrier where $U=E$.
The potential is repulsive for $\rho<r$,
zero for $\rho=r$
and attractive for $\rho>r$.
Note that the inverse function $r(\rho)$ may be multi-valued,
as shown in Fig.\ \ref{fig:rhohide}A. 
The additional values of $r(\rho)$ describe the turning
points of additional bound trajectories 
for the same energy $E$ and the angular momentum
that corresponds to the impact parameter $b$.
Scattering does not probe such bound states,
although the trajectories of scattered rays 
may enter the same region
for different impact parameters $b$.
As we show, the possibility of such elusive bound
trajectories indicates ambiguities in scattering.

In the following, 
we express the description of scattering in central potentials
as a tomographic projection for the integrated scattering angle 
\begin{equation}
\phi = \int_\infty^b \chi\, db \,.
\label{eq:phi}
\end{equation}
First, we represent Eq.\ (\ref{eq:chi}) as 
\begin{eqnarray}
\chi &=& 2b\left(
\int_{b}^\infty
\frac{d\rho}{\rho\sqrt{\rho^2-b^2}} -
\int_{r_0}^\infty
\frac{dr}{r\sqrt{\rho^2-b^2}} 
\right)
\nonumber\\
&=& 2b \int_{b}^\infty \frac{W'}{a}\, d\rho 
\label{eq:chi1}
\end{eqnarray}
in terms of 
\begin{equation}
W = \ln (\rho/r) \,, \quad a = \sqrt{\rho^2-b^2} \,.
\label{eq:w}
\end{equation}
A prime indicates 
differentiation with respect to the turning parameter.
We call $W$ scattering function.
When $r(\rho)$ is multi-valued the integration contour 
is understood to follow accordingly.
Since 
\begin{equation}
\frac{d}{db}\int_{b}^\infty W' a\,d\rho 
+ \int_{b}^\infty \frac{W'b}{a}\, d\rho =
\left.-W'a\right|_{\rho=b} = 0 \,,
\end{equation}
we obtain for the integrated scattering angle 
\begin{equation}
\phi = -2\int_{b}^\infty W' a\,d\rho  = 
2\int_{b}^\infty W a'\,d\rho =
\int_{-\infty}^{+\infty} W da
\,.
\label{eq:result}
\end{equation}
This result has a simple geometrical meaning
illustrated in Fig.\ \ref{fig:w}:
\begin{figure}[t]
\vspace*{-1mm}
\begin{center}
\includegraphics[width=20.0pc]{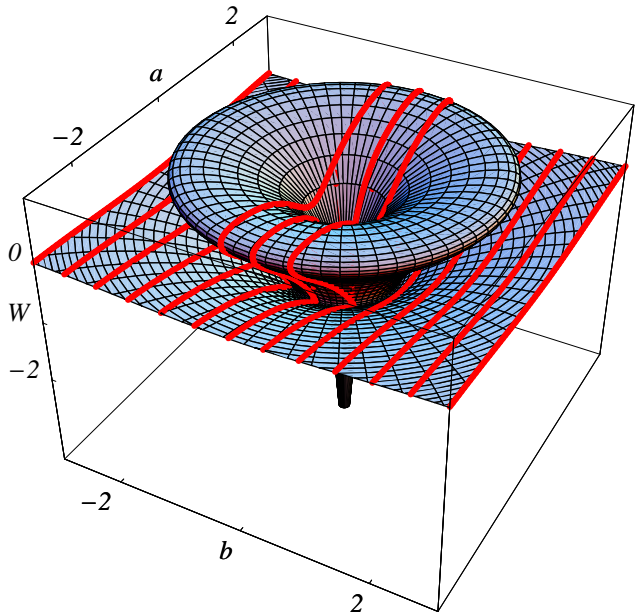}
\vspace*{-2mm}
\caption{\label{fig:w}
{\tiny
Scattering tomography.
The integrated scattering angle $\phi$ is a projection
of the scattering function $W$ 
onto the impact parameter $b$ along the 
red lines of the fictitious parameter $a$.
The radius $\rho$ in this auxiliary $(a,b)$ plane 
is the turning parameter (\ref{eq:rho}).
The scattering function, defined in Eq.\ (\ref{eq:w})
and obtained from Fig.\ (\ref{fig:rhohide}),
may be multi-valued, as shown here.
The folds of $W$ visualize the ambiguities 
of scattering.
}
\vspace*{-1mm}
}
\end{center}
\end{figure}
imagine that $a$ and $b$ constitute a plane
of impact parameters where one, $b$, is
experimentally accessible and the other, $a$, is not.
The scattering function $W$ depends only on the radius
$\rho=\sqrt{a^2+b^2}$,
both directly by definition (\ref{eq:w}) and in $r(\rho)$.
Equation (\ref{eq:result}) shows that
the integrated scattering angle is a projection of the 
rotationally symmetric object $W(\rho)$
onto the experimentally accessible impact parameter $b$
in exactly the same way as objects are projected in
classical tomography \cite{Tomo}
or Wigner functions in quantum tomography \cite{Leo,LeoJex}.
If $r(\rho)$ is single-valued,
one can invert the projection by 
the inverse Abel transformation \cite{Leo,LeoJex}
\begin{equation}
W = 
-\frac{1}{\pi}\int_{\rho}^\infty 
\frac{\chi\, db}{\sqrt{b^2-\rho^2}}
\,,
\label{eq:inverse}
\end{equation}
a special case 
of the inverse Radon transformation \cite{Leo}. 
If $r(\rho)$ is multi-valued one can hide features
of the potential in the folds of $W$, 
as Fig.\ \ref{fig:w} illustrates. 

\begin{figure}[t]
\vspace*{-3mm}
\begin{center}
\includegraphics[width=19.0pc]{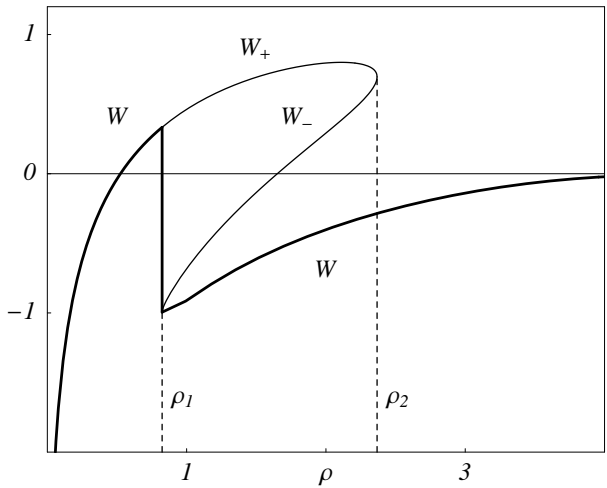}
\caption{\label{fig:fold}
{\tiny
Multi-valued scattering function $W$
obtained from Fig.\ (\ref{fig:rhohide})
according to the definition (\ref{eq:w}).
The $W(\rho)$ function is folded between 
the turning parameters $\rho_1$ and $\rho_2$
where 
$W_+$ denotes the top and 
$W_-$ the bottom curve of the fold.
}}
\end{center}
\vspace*{-8mm}
\end{figure}

Consider the scattering ambiguities 
where the scattering function $W$ is multi-valued.
The simplest case corresponds to a single fold in $W$
between two turning parameters $\rho_1$ and $\rho_2$,
as shown in Figs.\ \ref{fig:w} and \ref{fig:fold}.
We use the inverse Abel transformation (\ref{eq:inverse})
to construct a potential, described by $W_0(\rho)$,
that exhibits the same scattering characteristics as $W$.
Figure \ref{fig:w} indicates that
$W$ and $W_0$ agree for $\rho>\rho_1$,
because all projections lie under the fold.
For $\rho<\rho_1$ the scattering angle $\chi$ is,
according to Eq.\ (\ref{eq:chi}),
\begin{equation}
\chi = 2b\int_{b}^\infty
\frac{W' d\sigma}{\sqrt{\sigma^2-b^2}} +
2b\int_{\rho_1}^{\rho_2}
\frac{W'_+-W'_-}{\sqrt{\sigma^2-b^2}}
\,d\sigma
\,.
\end{equation}
where the integration variable $\sigma$ refers to
the turning parameter,
$W$ follows the solid curve in Fig.\ \ref{fig:fold},
with a jump at $\rho_1$,
whereas $W_+$ denote the top and 
$W_-$ the bottom curve of the fold.
Since the inverse Abel transformation (\ref{eq:inverse})
uniquely inverts the first term in $\chi$,
we obtain for the difference between $W$ and $W_0$
\begin{eqnarray}
W-W_0 &=& 
\frac{2}{\pi} \int_\rho^{\rho_1} \int_{\rho_1}^{\rho_2}
\frac{b\,(W'_+-W'_-)\,d\sigma\,db}
{\sqrt{(b^2-\rho^2)(\sigma^2-b^2)}}
\nonumber\\
&=& \frac{2}{\pi}  \int_{\rho_1}^{\rho_2}
(W'_+-W'_-)\, \arctan\left(
\frac{\sqrt{\rho_1^2-\rho^2}}{\sqrt{\sigma^2-\rho_1^2}}
\right) d\sigma
\nonumber\\
&=& \frac{2}{\pi}  \int_{\rho_1}^{\rho_2}
(W_+-W_-)\,
\frac{\sqrt{\rho_1^2-\rho^2}}{\sqrt{\sigma^2-\rho_1^2}}\,
\frac{\sigma \,d\sigma }{\sigma^2-\rho^2}
\label{eq:dw}
\end{eqnarray}
by partial integration, utilizing that the boundary term
vanishes, because $W_-(\rho_2)=W_+(\rho_2)$.
Since $W_+>W_-$  the ambiguous $W$ must 
exceed $W_0$ in the single-valued region inside $\rho_1$,
which implies that the radius 
$r=\rho\,\exp(-W_0)$ is greater than $\rho\,\exp(-W)$.
The fold of multi-valuedness thus
magnifies the scattering structure of the potential. 
In particular, 
for ambiguous scattering potentials,
the zero of $\rho(r)$ is closer to the origin
than for the equivalent non-ambiguous one.
Since this zero corresponds 
to the potential barrier beyond which one can hide,
nothing is gained, quite the opposite.
This feature continues in the general case of
several folds in $W$, because one could replace $W$
by equivalent single-valued $W_0$ with the same
scattering characteristics, starting from the outmost fold
and proceeding to the inside.

\begin{figure}[t]
\begin{center}
\includegraphics[width=19.0pc]{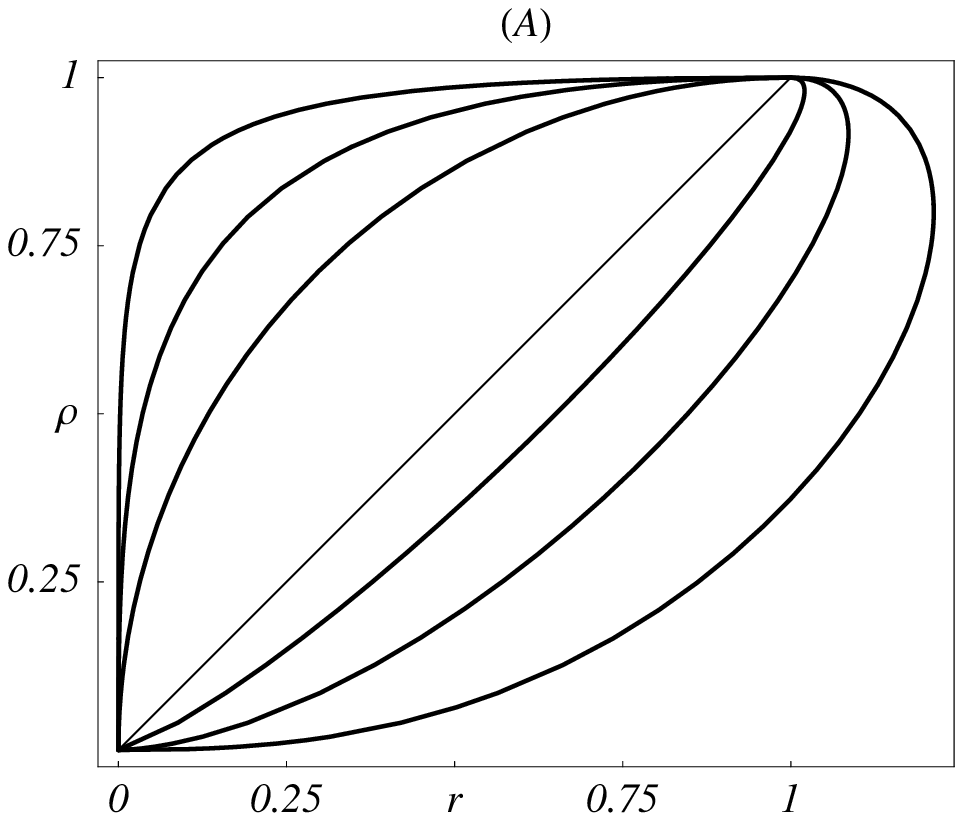}
\includegraphics[width=20.0pc]{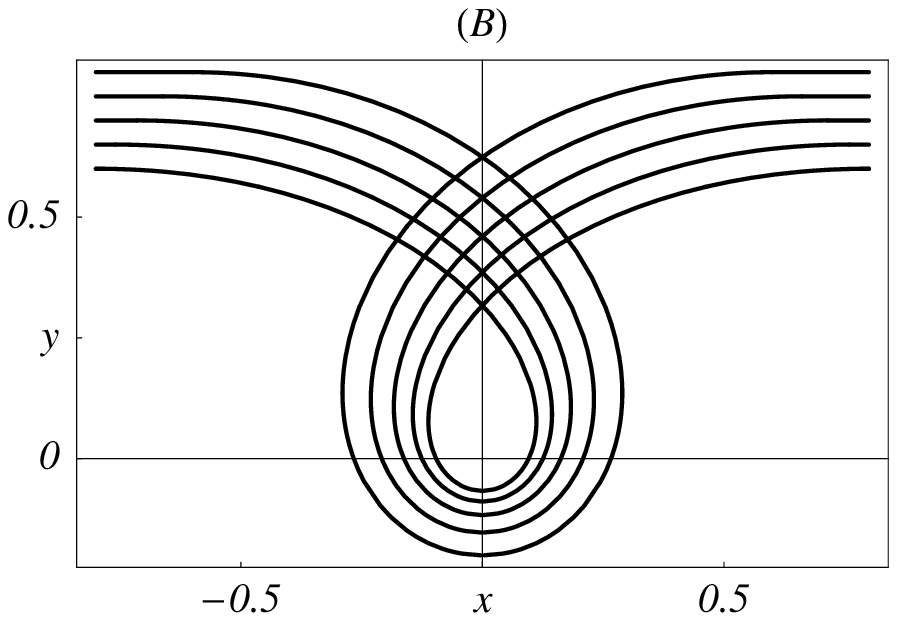}
\caption{\label{fig:rho}
{\tiny
Uniform bending.
Subfigure A shows the turning parameters
obtained from Eq. (\ref{eq:nu})
for the winding numbers
$\nu \in \{2,1,0.5,-0.1,-0.2,-0.3\}$ and $b_0=1$.
For negative $\nu$, $\rho(r)$ is multi-valued
and hence unphysical.
Subfigure B illustrates the uniform loops
of trajectories for $\nu=1$.
\vspace*{-8mm}
}}
\end{center}
\end{figure}

An alternative way of hiding the presence of a potential 
would be to let the trajectories leave at
scattering angles that are multiples of $2\pi$,
{\it i.e} to turn them around in precisely adjusted loops.
Suppose that for impact parameters $b$ smaller
than a critical $b_0$ the trajectories are uniformely turned by
$\chi=-2\pi\nu$ and are not affected for $b$ larger than $b_0$. 
Here $\nu$ may be a real number, not only an integer,
for the sake of generality.
Assuming that $r(\rho)$ is single-valued,
we obtain from the inverse Abel transformation 
(\ref{eq:inverse})
\begin{equation}
r = \left\{
\begin{array}{lcl}
\rho \left({b_0}/{\rho} + 
\sqrt{{b_0^2}/{\rho^2}-1}\right)^{-2\nu}
&:& \rho < b_0 \\
\rho &:& \rho \ge b_0
\end{array}
\right.
\,.
\label{eq:nu}
\end{equation}
Figure \ref{fig:rho} illustrates the curves of $\rho(r)$.
Clearly, $r(\rho)$ is single-valued by definition.
For  $\nu<0$ the potential would be repulsive,
because the trajectories are deflected,
but in this case the function $\rho(r)$
itself is multivalued.
Consequently, 
no central potential exists that  uniformly deflects trajectories.
For $\nu>0$ the potential is attractive,
as one would expect to be necessary for bending
trajectories around the center of force.
The case $\nu=1/2$ corresponds to 
a Kepler potential \cite{LL1} or
the Eaton lens \cite{KerkerScattering}
developed in radar technology.
In the limit $\rho\rightarrow 0$ 
we get from Eq.\ (\ref{eq:nu}) the asymptotics
$\rho/r \sim (2/r)^{2\nu/(2\nu+1)}$, and hence,
according to Eq.\ (\ref{eq:rho})
the potential $U$ diverges 
with the power $-4\nu/(2\nu+1)$ for small $r$.
One cannot hide anything here.
In the limit $\nu\rightarrow\infty$
of infinitely many cycles
$U$ approaches near the origin
the $1/r^2$ potential 
of fatal attraction \cite{LL1}.
Figure \ref{fig:rho}B illustrates the case where the
trajectories are turned around by $2\pi$. 

Although one cannot directly apply 
the ambiguous scattering of
isotropic
and centrally symmetric media 
to construct an invisibility device,
one can use their singularities to improve 
anisotropic devices.
Such a device is designed to facilitate a coordinate
transformation with a hole \cite{Pendry}.
Anything inside the hole is hidden by construction \cite{Pendry}.
Consider a two-dimensional case in polar coordinates.
Suppose that the radius $r$ is mapped onto $r'$ such that
$r'$ reaches the radius of the hole at $r=0$ as
\begin{equation}
\frac{\partial r'}{\partial r}\sim \alpha r^{-s}
\quad\mbox{for}\quad r \sim 0 \,,
\label{eq:trans}
\end{equation}
where $\alpha$ and $s$ are non-negative constants.
Beyond the outer radius $b_0$ of the cloak the 
coordinates $r'$ shall coincide with $r$.
Assume in unprimed space
the isotropic and radially symmetric refractive-index 
profile $n(r)$ with perfect impedance matching.
Reference \cite{Pendry} gives a recipe to calculate
the dielectric $\varepsilon$ and magnetic $\mu$
that facilitates the coordinate transformation (\ref{eq:trans}).
We find
\begin{equation}
\varepsilon_r'=\mu_r' \sim \frac{\alpha r^{1-s}}{r'}\,n(r) \,,\quad
\varepsilon_\varphi'=\mu_\varphi'
\sim \frac{r^{s-1}}{\alpha r'}\,n(r) \,.
\end{equation}
Suppose that we use a profile where $n^2/2$
corresponds \cite{LeoNotes} to the
$E-U$ of uniform bending (\ref{eq:nu}) with the definition (\ref{eq:rho})
and $\nu=1$.
If we choose $s=1/3$ the
singularity of $U$ compensates for the zero in the
refractive index in real space that would otherwise
imply \cite{Pendry} 
that the speed of light tends to infinity at the inner surface of  the cloak.
The phase velocity in radial direction is finite.
On the other hand, the speed of light in angular direction
tends to zero with the power $4/3$.
Our simple example
indicates that invisibility devices with finite phase velocity
are possible in principle.
In our case,
wrapping light around the invisibility device
stratifies the optical wavefronts.
However,  there is a price to pay:
light propagation with finite phase velocity
around an object 
inevitably causes time delays
that result in wavefront dislocations at the boundary
\cite{LeoNotes}.
The invisibility is perfect for rays,
but not for waves.

{\it Conclusions.---}
Scattering in central potentials corresponds
to a tomographic projection
that visualizes scattering ambiguities.
Such ambiguities are limited, though:
central potentials are not suitable
to achieve the same
scattering characteristics as empty space.
Therefore, highly asymmetric refractive-index profiles 
\cite{LeoConform,LeoNotes}
or highly anisotropic media \cite{Pendry}
are required to design invisibility devices,
which, interestingly, can operate with 
a finite speed of light.
Otherwise, trying to hide things uniformly from all sides
just magnifies them.

The paper was supported by 
the Alexander von Humboldt Foundation,
the Leverhulme Trust and 
King Saud University.

\end{document}